\documentclass[pra,showpacs,showkeys,preprintnumbers,nofootinbib,twocolumn,sort&compress]{revtex4}
\usepackage{amsmath,amsfonts,amssymb,amsthm}
\usepackage{bm}
\def\be{\begin{equation}}
\def\ee{\end{equation}}
\def\ba{\begin{eqnarray}}
\def\ea{\end{eqnarray}}

\def\f#1#2{\frac{#1}{#2}}
\def\te#1{\text{#1}}
\def\tilde#1{\widetilde{#1}}

\def\ket#1{\left| #1 \right\rangle}

\def\lrpBig#1{\Big( #1 \Big)}

\def\({\left(}
\def\){\right)}
\def\<{\left\langle}
\def\>{\right\rangle}

\def\Tr{{\rm Tr}}
\def\l:{\mathopen{:}\,}
\def\r:{\,\mathclose{:}}

\def\R{\mathbb{R}}

\def\C{\mathbb{C}}

\def\m{\mu}

\def\brho{{\bm \rho}}

\def\al{\alpha}

\def\g{\mathfrak{g}}
\def\fN{\mathfrak{n}}

\def\H{\mathcal{H}}
\def\Ha{\H_A}
\def\Hb{\H_B}

\def\cM{\mathcal{M}}

\def\cH{\mathcal{H}}

\def\cD{\mathcal{D}}
\def\cL{\mathcal{L}}

\def\cA{\mathcal{A}}

\def\cE{\mathcal{E}}

\def\I{\bm{1}}

\def\bel#1{\begin{equation} \label{#1}}
\def\ee{\end{equation}}
\def\bea{\begin{eqnarray*}}
\def\eea{\end{eqnarray*}}

\def\er{\eqref}

\def\lrp#1{\left( #1 \right)}
\def\Superop{\cE}
\newcommand{\til}{\widetilde}

\def\su{\mathfrak{su}}
\def\gl{\mathfrak{gl}}

\def\ep{\epsilon}
\def\h{\mathfrak{h}}

\def\ff{\mathfrak{f}}

\def\Im{\operatorname{Im}}
\def\End{\operatorname{End}}

\def\tr{\operatorname{tr}}

\def\ad{\mathrm{ad}}

\newtheorem{definition}{Definition}
\newtheorem{theorem}{Theorem}

\newtheorem{lemma}{Lemma}

\newtheorem{assumption}{Assumption}
\theoremstyle{remark}

\begin{document}

\title{Lie Algebras and Suppression of Decoherence in Open Quantum Systems}

\author{William Gordon \surname{Ritter}}
\affiliation{Harvard University Department of Physics \\
    17 Oxford St., Cambridge, MA 02138}

\date{April 15, 2005}

\begin{abstract}
Since there are many examples in which no decoherence-free
subsystems exist (among them all cases where the error generators
act irreducibly on the system Hilbert space), it is of interest to
search for novel mechanisms which suppress decoherence in these
more general cases. Drawing on recent work (quant-ph/0502153) we
present three results which indicate decoherence suppression
without the need for noiseless subsystems. There is a certain
trade-off; our results do not necessarily apply to an arbitrary
initial density matrix, or for completely generic noise
parameters. On the other hand, our computational methods are novel
and the result---suppression of decoherence in the error-algebra
approach \emph{without} noiseless subsystems---is an interesting
new direction.
\end{abstract}

\pacs{03.65.Yz, 03.67.Pp, 03.65.Fd, 02.20.Qs, 02.20.Sv}

\keywords{decoherence, quantum channels, open systems, Markovian
dynamics}

% \classification{Mathematics Subject Classification}{81T08}

\maketitle

\section{Introduction}

The central obstacle in the experimental realization of quantum
computers has proven to be maintaining the quantum coherence of
states \cite{NielsenChuang}. The main cause of this degradation of
the quantum coherence is the coupling of the computer to the
environment, and the decoherence induced by this coupling.

A variety of schemes for protecting quantum information have been
developed, including quantum error correction codes
\cite{Bennett,Gottesman,Knill:97a,Calderbank}, decoherence free
subspaces
\cite{Zanardi:97a,Zanardi:97c,Zanardi:98a,Lidar:PRL98,Lidar:1998nd},
noiseless subsystems \cite{Knill:99a}, bang-bang decoupling
\cite{Viola:1998a}, and topological quantum computation
\cite{TOP}. The first four of these techniques are closely related
to each other and can be described in a simple unified framework
based on representations of the algebra of errors
\cite{Knill:99a,Zanardi:99d,RZ:2000}. More recently, \citet*{ZL}
showed that topological quantum computation also falls into the
error-algebra framework.

Our approach, like those mentioned above, uses the (Lie algebraic)
structure of the physical decoherence process itself to protect
quantum coherence. We start from the assumption that the Lindblad
operators generate a representation of a Lie algebra, and show
that if this representation is irreducible, then decoherence-free
subsystems and subspaces will not exist. Thus if one is to have
any hope of controlling decoherence in the irreducible case, some
other mechanism is needed.

\subsection{Markovian Dynamics}

The dynamics of a quantum system $A$ coupled to a heat bath $B$,
which together form a closed system, is described by a Hamiltonian
\cite{book:OQS}:
\[
    (H_A \otimes \I_B) \  + \ (\I_A \otimes H_B)
    \ +\  H_I,
\]
where $H_A, H_B,$ and $H_I$ are the system, bath, and interaction
Hamiltonians respectively. The \emph{Markovian Master equation}
\cite{Lindblad:76a,IQD} (also ``semigroup master equation'' or
SME) provides the most general form for time evolution of the
system density operator $\rho = \rho_A$, which acts on a Hilbert
space $\Ha$:
\bel{master}
    \f{d\rho}{d t} = - \f{i}{\hbar} [H_A, \rho] + \cL(\rho) \, .
\ee
Here $\cL$ is called the Lindbladian or \emph{dissipator}, and
takes the general form
\begin{eqnarray}
    \label{lin}
    \cL(\rho) &=& \f12 \sum_{\al,\beta}
    a_{\al,\beta} L_{F_\al, F_\beta}(\rho),
    \\
    L_{F_\al, F_\beta}(\rho)
    &\equiv&
        [F_\al, \rho F_\beta^\dag ] + [F_\al\rho, F_\beta^\dag] \,
        .
        \label{lin2}
\end{eqnarray}
The equation of motion remains invariant under an arbitrary
unitary transformation of the Lindblad operators. The special case
corresponding to $\cL(\rho) = 0$ is the \emph{von Neumann
equation}, and describes unitary evolution. Therefore in general,
$\cL(\rho)$ contains precisely those terms in the evolution
equation which are responsible for decoherence. The derivation of
eqns.~\er{lin} and \er{lin2} from fundamental assumptions is done
in many places; \citet{Bacon:1999vr} have a particularly nice
treatment.

The coefficient matrix $(a_{\al,\beta})$ is assumed to be
time-independent and Hermitian. By diagonalizing $a$, it follows
that \er{lin} may also be written in ``diagonal standard form,''
replacing the $F_\alpha$ operators by suitable linear combinations
$V_a$. This yields
\bel{diagonalform}
    \cL(\rho) = \f12 \sum_a L_{V_a,V_a}(\rho) \, .
\ee
The noise coefficients $a_{\al,\beta}$ have been absorbed within
the operators $V_a$. From \er{diagonalform}, using the relation
\mbox{$L_{A,A} = [A, [\rho,A]]$} which holds for a Hermitian
operator $A$, we find
\begin{align*}
    \cL(\rho) &= \f12 \sum_a [V_a, [\rho,V_a]]
    =
    \sum_a \lrp{ V_a \rho V_a - \f12 \{ \rho, V_a^2 \} }
    \cr
    &=
    \sum_a V_a \rho V_a - \f12 \Big\{ \rho, \sum_a V_a^2 \Big\} \, .
\end{align*}
If $\sum_a V_a^2 = c \I$ where $c \in \R$ is a constant, then we
have
\[
    \cL(\rho) = \sum_a V_a \rho V_a - c \rho \, .
\]
The first term is reminiscent of a quantum channel with Hermitian
Kraus operators. Indeed, an equivalent form of the Markov master
equation is
\[
    \f{d}{dt} \rho_t
    =
    -i [H, \rho_t]
    +
    \Phi[\rho_t] - \f12 \{ \Phi^*(\I), \rho_t \}
\]
where $\Phi$ is a completely positive map
\cite[p.~18,\,II.5.5]{AlickiLendi}.

In applications to real-world systems, the matrix elements
$a_{\alpha \beta}$ contain physical parameters such as lifetimes,
longitudinal or transverse relaxation times, stationary
magnetization, etc. We will show that particularly simple dynamics
emerge if the coefficient matrix is not only Hermitian but also
real (hence symmetric) or approximately so.

The $F_\al$ describe various decoherence processes, and for this
reason, they are often called \emph{error generators}. The $F_\al$
are determined implicitly by the interaction Hamiltonian
\[
    H_I = \sum_\al F_\al \otimes B_\al ,
\]
where $\{ B_\al \}$ is a collection of operators on $\Hb$, called
the \emph{heat bath operators}. Let $\gl(\Ha)$ denote the vector
space of all linear transformations on $\Ha$, not necessarily
invertible. This is the Lie algebra of the group $GL(n)$ where $n
= \dim(\Ha)$. We now state the main assumption of the paper.

\begin{assumption} \label{LieAlg}
There exists a Lie algebra $\g$ and a faithful representation
$\phi : \g \to \gl(\Ha)$ such that
\bel{rep1}
    F_\al = \phi(f_\al), \quad \al = 0 \ldots M
\ee
for some linearly independent set $\{ f_\al \} \subset \g$.
\end{assumption}

This assumption is also basic to many other studies of
decoherence; see \cite{Lidar:PRL98,Bacon:1999vr,Ritter} for a few.
The trace of any generator of any representation of a compact
simple Lie algebra is zero \cite[Thm.~8.9]{Georgi}. Further, we
are primarily interested in representations for which all the
generators can be chosen to be Hermitian. This is always possible
for compact Lie algebras \cite[Sec.~2.4]{Georgi}. We may therefore
take $F_\al$ to be traceless and Hermitian.

Decoherence processes in which the $F_\al$ are Hermitian are
mathematically much simpler than the most general process, and
this is often the case of interest to applications in quantum
computing. We recall for clarity the standard example of a quantum
computer made of $K$ qubits with $n = 2^K$ dimensional register
Hilbert space.

Each qubit has four possibilities: no error, or an error generated
by one of the three Pauli matrices. This means that each qubit
independently undergoes the action of the standard depolarizing
channel
\bel{sdc}
    \rho \to \sum_\mu M_\m \rho M_\m
\ee
with Kraus operators
\bel{su2}
    M_0 = \sqrt{1-p} \ \I, \quad
    M_i =  \sqrt{\f{p}{Z}}\ \sigma_i \quad\  (i = 1 \ldots 3) \, .
\ee
where $\sigma_i$ are Pauli matrices and $Z$ is a normalization
constant equal to 3/4 for the spin-$1/2$ representation.

The maximum possible complexity of error generation is when
combined errors from any number of qubits are generated. The
$4^K-1$ error generators $F_\al$ for such a process are basis
elements of $\g = \su(n) = \su(2^K)$ in the defining
($n$-dimensional) representation. As the defining representation
is irreducible, there are no DF subspaces. We have written
\er{su2} in a notation compatible with the generalization to
arbitrary Lie algebras done in Section
\ref{sec:representation-theory} and more completely in
\cite{Ritter}.

\subsection{Irreducible Representations Have No Decoherence-Free Subsystems}

\citet*{Knill:99a} discovered a method for decoherence-free coding
into {\em subsystems} instead of into subspaces which has since
received much attention; see
\cite{Knill:99a,Zanardi:99d,DeFilippo:99a}. In this section, we
remark that in the situation of Assumption \ref{LieAlg}, existence
of a DF subsystem implies that the representation is reducible.

The definition of a DF subsystem begins with the obvious statement
that if the $F_\al$ are Hermitian and one of them is the identity,
then the associative algebra $\cA$ they generate is unital and
closed under adjoints. Therefore, $C^*$-algebra methods may be
applied.

Let $\cM(d,\C)$ denote the space of $d\times d$ complex matrices.
Any basis for $\cH$ determines a matrix representation $\phi : \cA
\to \cM(d,\C)$, where $d = \dim(\cH)$, simply by expressing each
operator in this basis. The idea is that $\phi$ may be
\emph{repetitive}, i.e. it may happen that {\small
\[
    \phi(a) =
    \begin{pmatrix}
    { \left.
    \begin{matrix}
    \phi_{d_1}(a) & & 0 \\ \ & \ddots & \ \\ 0 & \ & \phi_{d_1}(a)
    \end{matrix}
    \right] n_1   }
    \ & \  \\
    \ &
    \left.
    \begin{matrix}
    \phi_{d_2}(a) & & 0 \\ \ & \ddots & \ \\ 0 & \ & \phi_{d_2}(a)
    \end{matrix}
    \right] n_2 \\
    \ & \ddots
    \end{pmatrix}
\] }
where $\phi_{d_i} : \cA \to \cM(d_i,\C)$. The $n_1, n_2, \ldots$
label the sizes of the corresponding sub-matrices. In fancier
notation, $\phi(a) = \sum_i \I_{n_i} \otimes \phi_{d_i}(a)$. This
is also called the \emph{central decomposition}.

If this happens, then we may take a vector $\Psi \in \cH$ which is
also repetitive, so that the first $d_1$ components of $\Psi$ take
the form $\al v$, where $\al \in \C, v \in \C^{d_1}$, and the
second $d_1$ components take the form $\beta v$ for the same $v$,
etc. We may do this $n_1$ times. Then $\phi(a)\Psi=(\alpha w,
\beta w, \ldots)$ where $w = \phi_{d_1}(a)v$. The coefficient vector
$(\al, \beta, \ldots) \in \C^{n_1}$ can just as well be extracted
after applying the operator $\phi(a)$ as before. In other words,
the information contained in the $\alpha, \beta, \ldots$ is
protected under this decoherence process.

\begin{theorem} \label{thm:nodfss}
Decoherence-free subsystems do not exist if the error operators
$F_\al$ generate an irreducible representation.
\end{theorem}

\begin{proof}
If the $F_\al$ generate an irreducible matrix representation of a
semisimple Lie algebra $\g$, then $\cA$ will be the full algebra
of Hermitian operators over $\cH$. Therefore, the homomorphism
$\phi : \cA \to \cM(d,\C)$, where $d = \dim(\cH)$ is not
repetitive. In the above notation, $n_1 = 1$, there is no $n_2$
and $d_1 = \dim(\cH)$. Second proof: noiseless degrees of freedom
are associated with observables in ${\cal A}^\prime$, the
commutant algebra. By Schur's Lemma, $\cA'$ is the trivial algebra
if $\phi$ is irreducible.
\end{proof}

This theorem implies the corresponding result for DF subspaces as
a special case. Conditions under which DF subspaces can exist have
previously been studied in both the Lindblad (Markovian)
formulation \cite{Zanardi:98a,Lidar:PRL98} and for the
non-Markovian case \cite{Zanardi:97a}. Theorem \ref{thm:nodfss}
extends their results to noiseless subsystems.

\subsection{Irreducible Representations Return to Equilibrium}

In this section we prove a property of the time-evolution of open
quantum systems defined by irreducible representations. This
property is not used directly in the rest of the paper, but it
holds independent interest. The thermal Gibbs state
\[
    \rho_\beta = \f{e^{-\beta H}}{\tr e^{-\beta H}}
\]
is a stationary state for the Markovian dynamics of a system
coupled to a heat bath. The question then arises: under what
conditions does the system return to equilibrium for an initial
state $\rho$?

\begin{theorem}
Suppose that a Markovian system coupled to a heat bath is
described by Lindbladian \er{lin}--\er{lin2}, with $F$-operators
that form a set of generators for a nontrivial $n$-dimensional
representation of a semisimple Lie algebra. The system returns to
thermal equilibrium if and only if the representation is
irreducible.
\end{theorem}

\begin{proof}
Return to equilibrium for arbitrary initial state $\rho$ will
happen iff
\bel{returnto-eq}
    \lim_{t \to \infty} e^{\bm{L} t} \rho = \rho_\beta
    \ \te{ for all } \ \rho \in \cD(\cH),
\ee
where $\bm{L} = - i/\hbar [H_A, \rho] + \cL(\rho)$ denotes the
right side of \er{master}, and $\cD(\cH)$ denotes the space of
density operators on $\cH$. By assumption, the $F$-operators are
Hermitian, in which case \er{returnto-eq} is equivalent to the
statement that any operator $X$ commuting with all the $V_a$ must
look like $X =c\,\I$ for some constant $c$. This is equivalent by
Schur's lemma to the statement that the $V_a$ generate an
\emph{irreducible} representation.
\end{proof}

\subsection{Models of Decoherence}

\citet{Lidar:PRL98} classified the subspaces in which there is
\emph{generically} no decoherence in this model, where ``generic''
means that the decoherence is suppressed independently of the
noise parameters (i.e. the $a$-matrix of \er{lin}) and of the
initial conditions. If $\g$ is \emph{semisimple} (this assumption
includes the Lie algebras of $SU(n), SO(n), Sp(2n), G_2, F_4,
E_6,E_7,E_8$, and all direct sums of such algebras), then
\citet{Lidar:PRL98} characterize the decoherence-free (DF)
subspaces as those which are annihilated by every one of the
representation matrices. At an intuitive level, this is not
surprising since on such a subspace, the Lindblad operator
$L_{F_\al, F_\beta}(\rho) = [F_\al, \rho F_\beta^\dag ] +
[F_\al\rho, F_\beta^\dag]$ is identically zero, and hence
time-evolution is governed by the von Neumann equation. As
representations of $\g$, DF subspaces are always composed of
singlets.

As discussed above, there are many interesting quantum systems
which do not possess decoherence-free subspaces. Therefore, it is
of interest to search for other effects which suppress decoherence
to complement the existing techniques. On the other hand, at the
highest level of generality, completely arbitrary decoherence is
allowed by quantum mechanics. Thus it seems necessary to exploit
symmetry in some way in order to achieve a suppression. Therefore
we search for Lie-algebraic conditions which suppress the
Lindbladian, but allowing some dependence on the noise parameters
and on initial conditions.

The noise parameters and initial conditions, while difficult to
experimentally fine-tune, are also not entirely out of the hands
of laboratory control. Further, because the solution operator
$e^{\bm{L} t}$ depends smoothly on any linear parameters contained
in $\bm{L}$, the Lindbladian $\cL(\rho)$ depends smoothly on the
parameters $a_{\al,\beta}$, and hence a tiny variation in
$a_{\al,\beta}$ results in a correspondingly small change in the
dynamics.

\section{Algebraic identities for Lindblad operators}

\subsection{Suppression of Decoherence Without Noiseless
Subsystems}

Before continuing, we record a set of simple algebraic identities
for Lindblad operators. The Lindblad operator \er{lin2} may be
studied generally in terms of its matrix-valued bilinear form
\bel{bilform}
    L_{A,B} = 2 A \rho B^\dag - \{ \rho, B^\dag A \} .
\ee
Suppose that $A,B$ are both Hermitian. We assert that
\begin{eqnarray}
    \label{LAB}
    L_{A,B} &=& 2[A,B]\rho + 2A [\rho, B] - B[\rho,A] - [\rho,B]A
    \qquad \\
    \label{AB+BA}
    L_{A,B} &+& L_{B,A} = [A, [\rho,B]] + [B, [\rho,A]]
    \\
    \label{LAA}
    L_{A,A} &=& [A, [\rho,A]] \, .
\end{eqnarray}
hence if $A$ commutes with $\rho$ then $L_{A,A}=0$. If $[\rho,B]$
commutes with $A$ then
\bel{rhoB}
    L_{A,B} = 2[A,B]\rho + A [\rho, B] - B[\rho,A] .
\ee
The latter formula is antisymmetric in $A,B$. If $[\rho,A]$ also
commutes with $B$ then \er{rhoB} holds with $A$ and $B$ switched,
hence $L_{A,B} = - L_{B,A}$. This also follows immediately from
\er{AB+BA}. If $[\rho, A] = A$ and $[\rho,B] = B$ then
\bel{rhoA=A}
    L_{A,B} = 2[A,B](\rho + \I),\ \te{ hence  }\
    L_{A,B} = - L_{B,A}.
\ee

We now investigate the consequences of Assumption \ref{LieAlg}. We
identify elements $x \in \g$ with the operators $\phi(x)$ acting
on $\cH$.

\begin{lemma}\label{lem:adj1}
Suppose that for each $x \in \g$, $[\rho,x]$ commutes with all of
$\g$ and $a_{\al\beta}$ is real. Then $\cL(\rho) = 0$.
\end{lemma}

\begin{proof} The assumption entails that $[\rho,F_\al]$ commutes
with $F_\beta$ for all $\al, \beta$. Then \er{AB+BA} implies that
$L_{F_\alpha,F_\beta} = -L_{F_\beta,F_\alpha}$ for all $\alpha,
\beta$, but the sum $\cL(\rho)$ involves these objects in symmetric
combinations.
\end{proof}

It is natural to ask: when can the conditions of Lemma
\ref{lem:adj1} be satisfied? If $\rho$ can be expressed as a
linear combination of elements of $\g$ then the condition reduces
to
\[
    \ad_\rho(\g) \subset Z(\g) ,
\]
where $Z(\g)$ is the center of $\g$.

Generally, it may not be possible to express $\rho$ as a linear
combination of (the identity and) generators of $\g$
\cite{Ritter}. The vector space dimension $\dim(\g)$ will often be
smaller than $n^2-1$, the geometric dimension of the space of
density operators. However, it is possible to equip $\g$ with an
embedding into a larger algebra $\h$, and note that the conditions
of Lemma \ref{lem:adj1} will be satisfied if $\ad_\rho(\g)$ is
contained in the centralizer of $\g$ within $\h$, i.e. the set of
all elements in $\h$ that commute with all elements in $\g$.

\begin{lemma} \label{lem:adj2}
Suppose that $\ad_\rho$ is the identity on $\g$ (i.e.
$[\rho,F_\al] = F_\al$ for all $\al$) and $a_{\al\beta}$ is real.
Then $\cL(\rho) = 0$.
\end{lemma}

Lemmas \ref{lem:adj1} and \ref{lem:adj2} point to the following
general principle: \emph{suppression of decoherence, for an
initial state $\rho$, is related to the adjoint action of the
density matrix $\rho$ on the Lie algebra $\g$ containing the error
generators.}

\subsection{Lindblad Operators for su${}_n$}

The above analysis leads to exactly calculable formulae for
Lindblad operators in the $\bm{n}$ representation of $\su(n)$. Let
$\{ X_j \}$ be the canonical set of generators for $\su(n)$ in
this representation, satisfying $\Tr(X_i X_j) = 2 \delta_{ij}$.
\citet{MacFarlane} give a very clear discussion of $\su(n)$
generators and the identities that they satisfy; we use the same
notation and conventions.

The $L_{F_\al,F_\beta}$ are sums of commutators, hence they are
traceless and may therefore be expressed as a linear combination
of the $X_i$. An arbitrary $n \times n$ density matrix $\brho_v$
may be written in the Bloch representation
\[
    \brho_v = \f1{n} \lrpBig{ \I + \sum_a v_a X_a }, \ \
    v \in \R^{n^2-1}
\]
and write $L_{ij}$ for $L_{X_i, X_j}$.

For calculating with Lie algebras, we find it extremely useful to
use a convention even lazier than the Einstein summation
convention to systematically not write sums and repeated indices,
as follows. Each $X_a$ is a matrix, but we never write the matrix
indices; $v_a$ is a vector with the same dimension so $v \cdot X =
\sum_a v_a X_a$ is a matrix. Extend this to all tensors, so that
$v \cdot f_b \cdot X \longrightarrow \sum_{a,c} v_a f_{abc} X_c$
etc.

The standard basis for the $\su(n)$ algebra satisfies the basic
identities, which are derived in \cite{MacFarlane}:
\begin{align*}
    [\brho_v, X_b] &= \f{2i}{n} v \cdot f_b \cdot X ,
    \quad\ \
    X_a X_b = \f{2}{n} \delta_{ab} \I + Q_{ab} \cdot X,
    \cr
    & \quad [X_a, X_b] = 2i f_{ab} \cdot X \, ,
\end{align*}
This makes explicit calculations easy. For example, if the terms
in the diagonal standard form \er{diagonalform} happen to be the
$\su(n)$ generators, then using \er{LAA} we find
\[
    L_{ii}
    =
    [ X_i, \ \f{2i}{n} v \cdot f_i \cdot X ]
    =
    \f{4}{n} \, v \cdot f_i \cdot f_i \cdot X \, .
\]
Using \cite[eq.~(2.12)]{MacFarlane}, we then have
\bel{eq:sun-deco}
    \sum_i L_{ii} = -4 (v \cdot X) = 4(\I - n \rho) \, .
\ee
Thus we have derived a very simple formula for the Lindbladian of
$\su(n)$-decoherence. It vanishes iff $\rho = \f{1}{n} \I$.

\section{Symmetry Breaking} \label{sec:symmetrybreaking}

Symmetry breaking in this context can be modeled as the
introduction of new error generators which do not belong to the
representation of $\g$ which gave the original symmetry. The
standard assumption is that the symmetry is broken perturbatively
by modification of the system-bath Hamiltonian:
\bel{SBpert}
    H_{AB} \to H_{AB} + \epsilon H_{I, \te{sb}} \, .
\ee
where $H_{I, \te{sb}}$ denotes a symmetry-breaking interaction
between the system and the bath, and $\ep \ll 1$. As discussed by
\citet{Lidar:PRL98}, the new terms added to the Lindbladian are of
the form
\bel{SB}
    \cL'(\rho) =
    \sum_{\al,p} \lrpBig{ \tilde a_{\al p} L_{F_\al, \ep G_p} + h.c.}
    + \sum_{p,q} \tilde b_{pq} L_{\ep G_p, \ep G_q}
\ee
where $h.c.$ denotes the hermitian conjugate. Clearly the terms
proportional to $L_{\ep G_p, \ep G_q}$ are $O(\ep^2)$ and can be
neglected in first-order perturbation theory, so we turn to the
analysis of the $O(\ep)$ terms in the present context.

\begin{definition}\label{def:center}
Let $\h$ be a Lie algebra with subspaces $\g,\ff$.

\noindent (a) Following standard textbooks, we define the {\bf
centralizer} of $\g$ in $\h$ by
\[
    Z_\h(\g) = \{ x \in \h : [x,g] = 0 \ \forall \ g \in \g \} \,
    .
\]

\noindent (b) Let $\phi : \h \to \End(V)$ be a representation of
$\h$ on a vector space $V$ and let $\rho$ be an operator on $V$.
We say that a Lie algebra homomorphism $A : \h \to \h$ {\bf
switches centralizers} for $\g,\ff$ if
\[
    A(\ff) \subset Z_\h(\g) \ \te{ and } \
    A(\g) \subset Z_\h(\ff) \, .
\]
We also say that $\rho$ switches centralizers if $A = \ad_\rho$
does.\footnote{While Def.~\ref{def:center}(a) is standard, I have
never seen Def.~\ref{def:center}(b) before.}
\end{definition}

Assume that $\h$ admits a faithful representation $\phi : \h \to
\gl(\cH)$ on the system Hilbert space. Since $\phi$ is faithful,
identify $\h$ with its image under $\phi$, which makes sense since
we are just interested in operators on $\cH$.

Let $\ff$ and $\g$ be the Lie algebras generated by the
$F$-operators and the $G$-operators respectively. If $\rho$
switches centralizers for $\ff$ and $\g$, then \er{AB+BA} implies
\[
    L_{F_\al, \ep G_p} = - L_{\ep G_p,F_\al}
\]
and therefore, the corresponding term in \er{SB} is proportional
to
\[
    {\til a}_{\al p} - {\til a}_{\al p}^* = 2i \Im(a_{\al p}).
\]
If the couplings ${\til a}_{\al p}$ are \emph{real}, then the
$O(\epsilon)$ symmetry breaking terms \er{SB} are zero, and the
symmetry is perturbatively stable (to first order). We have
proven:

\begin{theorem} \label{[G,F]}
Let $\ff$ and $\g$ be the Lie algebras generated by the
$F$-operators and the $G$-operators respectively. If $\rho$
switches centralizers for $\ff$ and $\g$, and $\til a_{\al p} \in
\R$ for all $\al,p$, then $\cL'(\rho) = 0$. If $\til a_{\al,p}$
has an imaginary part, then the $O(\ep)$ terms in $\cL'(\rho)$ are
proportional to $\Im(\til a_{\al,p})$.
\end{theorem}

Physically, these interactions are \emph{safe} for quantum
computing, because they do not contribute to decoherence. It is
therefore of interest to locate examples of density matrices which
switch centralizers for certain subalgebras.

The most obvious example is the following. Suppose that $\h = \ff
\oplus \g$ is a direct sum, so in an appropriate basis we can
write all of the representation matrices in Block diagonal form;
schematically,
\[
    \phi(\h) = \begin{pmatrix} \phi(\ff) & 0 \\ 0 & \phi(\g)
    \end{pmatrix} \, .
\]
We will now continue with our convention of identifying $\ff$ with
its image under the faithful representation $\phi$. Suppose that
$\rho$ is also Block diagonal in the same basis, so that $[\rho,
\ff] \subset \ff$ and $[\rho, \g] \subset \g$. Under these
conditions, the identity homomorphism switches centralizers, i.e.
$\ff \subset Z_\h(\g)$ and $\g \subset Z_\h(\ff)$. Therefore,
$\ad_\rho$ also switches centralizers. This, like any example with
a DF subsystem, is a reducible representation, but it is plausible
that centralizer-switching occurs for irreducible representations
as well.

It is fruitful to apply this analysis to the case of a
symmetry-breaking perturbation \er{SB}. Theorem \ref{[G,F]}
implies that if $\ff$ and $\g$ switch centralizers, then the
cross-terms, which are those proportional to $\ep L_{F_\al, G_p}$
in \er{SB}, are zero. These are also the only $O(\ep)$ terms, so
this kind of symmetry breaking is a purely $O(\ep^2)$ effect!

\section{The Effect of Coarse-graining}
\label{sec:representation-theory}

\subsection{Lie Algebra Channels}

Quantum channels defined by Lie algebra representations were
defined and studied extensively in a separate paper
\citep{Ritter}. Here we recall the basic definitions and set
notation, before discussing the relationship to Markovian
dynamics.

\begin{definition} \label{def:gdc}
Let $\g$ denote a Lie algebra of dimension $k$, and let $\{ X_i :
i = 1, \ldots, k \}$ be a basis of $\g$. Let $\phi$ be a unitary
representation of $\g$ on a Hilbert space $\cH$. The
\emph{generalized depolarizing channel} or \emph{Lie algebra
channel} is defined to be the channel in which an error occurs
conditionally with probability $p$, causing an initial state
$\ket{\psi} \in \cH$ to evolve into an ensemble of the $k$ states
$\phi(X_i) \ket{\psi}$, all with equal likelihood.
\end{definition}

The Kraus operators for the channel of Definition \ref{def:gdc}
are given by
\bel{theMs}
    M_0 = \sqrt{1-p} \ \I, \quad
    M_i =  \sqrt{\f{p}{Z}}\ \phi(X_i) \, .
\ee
where $Z$ is a normalization constant, and will be determined
momentarily. The operators $M_\mu$ are hermitian if the
representation is unitary and if $p \in [0,1]$.

Let $K$ be the Killing form. If $\phi$ is irreducible and the
chosen basis satisfies pseudo-orthonormality
\bel{pseudo}
    K(X_i, X_j) = \fN\, \delta_{ij} \, , \quad
    \fN \ne 0,
\ee
then $\sum_i \phi(X_i)^2 = Z\cdot \I$ with $Z = \fN \, c_2(\phi)$
where $c_2(\phi)$ is the quadratic Casimir, and therefore the
condition of probability conservation $M_\mu M^\mu = \I$ is
satisfied. If $\phi$ is irreducible but the basis does not satisfy
\er{pseudo}, then Def.~\ref{def:gdc} does not define a channel. If
$\phi$ is reducible, then one must replace $Z \to \sum_p Z_p$
where the $Z_p$ are the normalization constants for the
independent irreducible components.

Using \er{theMs}, the Lie algebra channel has the explicit Kraus
decomposition
\bel{superop}
    \brho \to \Superop(\brho) = (1-p) \brho + \f{p}{Z}
    \sum_{i=1}^k \phi(X_i) \,\brho\, \phi(X_i)
    \, .
\ee
In what follows, we will write $X_i$ for $\phi(X_i)$ since the
distinction is clear from context.

The companion paper \citep{Ritter} gave methods for calculating
with these channels using Lie algebra identities. Let
$\Superop_{\g}$ denote the channel based on the fundamental
representation of $\g$; for example, $\Superop_{\su(n)}$ is based
on the $\bm{n}$ of $\su(n)$. The action of $\Superop_{\su(n)}$ on
a general hermitian matrix $\brho = n^{-1} (\tr(\brho) \I + v
\cdot X)$ can be calculated exactly in closed form, yielding
\bel{su_n-channel}
    \Superop(\brho_v)
    =
    \f1{n} \lrp{\tr(\brho) \I + \f{(1-p) n^2-1}{n^2-1}\, v \cdot X}
    \, .
\ee
In the qubit case, the coefficient of $v$ is $1-4p/3$, which is
exactly consistent with the standard qubit depolarizing channel
\er{sdc}-\er{su2}.

\subsection{Coarse-graining}

Markovian dynamics may be derived from quantum theory of
measurement by coarse-graining time. Expanding Lindblad's equation
$\dot{\rho} = - \frac{1}{2} \sum_\al a_\al([F_\al,\rho F_\al^\dag]
+ [F_\al \rho, F_\al^\dag])$ to first order in the short time
interval $\tau$ yields
%\bel{t+tau}
\[
    \rho(t+\tau )
    =
    M_0 \rho(t)M_0
        +
        \tau \sum_{\al} F_\al\rho(t)F_\al^\dag\, ,
\]
%\ee
with
\[
    M_0 = \I - \f{\tau}{2} \sum_i F_i^\dag F_i \, .
\]
Suppose that the Lindblad operators equal to properly normalized
Lie algebra generators, $F_i = Z^{-1/2} X_i$. Then
\[
    M_0 = \lrpBig{1-\f{\tau}{2}} \I \sim \sqrt{1-\tau}\ \I + O(\tau^2) .
\]
Also defining
\[
    M_i = \sqrt{\frac{\tau}{Z}} \, X_i
\]
shows that the coarse-grained Lindblad's equation is completely
equivalent to the Lie algebra channel \er{theMs} to first order in
$\tau$, under the mapping $\tau \to p$.

It is interesting that in the Lie algebra channel, $p$ is
probability of error while in Markovian dynamics, $\tau$ is the
coarse-graining time. In fact, their equivalence to first order
coincides with one of the defining properties of a Poisson
process: that the probability of exactly one change in a
sufficiently small time interval of length $\tau$ is proportional
to $\tau$, with constant of proportionality given by the event
rate for the generation of errors. This assumption is often made
(either implicitly or explicitly) in the theory of quantum error
correction and describes a \emph{quantum Poisson process}. An
example of a quantum Poisson process is spontaneous emission
\cite{Ahn}. For such processes, we expect $p \propto \tau$ to
first order in $p$ or $\tau$.

If we specialize to $\g = \su(n)$, then the evolution of the
channel was found to be \er{su_n-channel} with $p$ replaced by
$\tau$. Since \er{su_n-channel} is already linear in $p$, any
$O(\tau^2)$ terms just come along for the ride. The conclusion is
that in $\su(n)$ Markovian dynamics,
\bel{eq:sun-dynamics}
    \rho(t + \tau) \sim
    \f{\tr \rho(t)}{n} \, \I + \f{1}{n} \lrpBig{1 - \f{\tau n^2}{n^2-1}} \, v(t) \cdot X
    \, .
\ee
Eqn.~\er{eq:sun-dynamics} is something which, \emph{a priori} we
did not expect: an explicit formula, valid to first order in the
coarse-graining time, for the nontrivial Markovian system defined
by $\su_n$. The simplification here is related to the
simplification \er{eq:sun-deco} of the Lindbladian.

\section{Conclusions}

Since there are many examples in which no decoherence-free
subsystems exist (among them all cases where the error generators
act irreducibly on the system Hilbert space), it is of interest to
search for novel mechanisms which suppress decoherence in these
more general cases. We presented three results (Lemmas
\ref{lem:adj1} and \ref{lem:adj2}, Theorem \ref{[G,F]}) which
indicate decoherence suppression without the need for noiseless
subsystems. There is a certain trade-off; our methods do not
necessarily apply to an arbitrary initial density matrix, or for
completely generic noise parameters $a_{\al\beta}$. On the other
hand, our computational methods are novel and the
result--suppression of decoherence in the error-algebra approach
\emph{without} noiseless subsystems--is an interesting new
direction which warrants further study.

\bibliographystyle{apsrev}

\end{document}